\documentclass[prb, reprint, superscriptaddress, floatfix]{revtex4-2}

\usepackage{siunitx}
\usepackage{graphicx}

\begin{document}

\title{Long spin coherence times of nitrogen vacancy centers in milled nanodiamonds}

\author{B. D. Wood}
\email[]{ben.d.wood@warwick.ac.uk}
\affiliation{Department of Physics, University of Warwick, Coventry, CV4 7AL, United Kingdom}
\author{G. A. Stimpson}
\affiliation{Department of Physics, University of Warwick, Coventry, CV4 7AL, United Kingdom}
\affiliation{Diamond Science and Technology Centre for Doctoral Training, University of Warwick, Coventry, CV4 7AL, United Kingdom}
\author{J. E. March}
\author{Y. N. D. Lekhai}
\author{C. J. Stephen}

\author{B. L. Green}
\author{A. C. Frangeskou}
\altaffiliation[Current address: ]{Lightbox Jewelry, Orion House, 5 Upper St. Martins Lane, London, WC2H 9EA, United Kingdom}
\affiliation{Department of Physics, University of Warwick, Coventry, CV4 7AL, United Kingdom}
\author{L. Gin\'es}
\author{S. Mandal}
\author{O. A. Williams}
\affiliation{School of Physics and Astronomy, Cardiff University, Queen’s Building, The Parade, Cardiff,
CF24 3AA, United Kingdom}
\author{G. W. Morley}
\email[]{gavin.morley@warwick.ac.uk}
\affiliation{Department of Physics, University of Warwick, Coventry, CV4 7AL, United Kingdom}
\affiliation{Diamond Science and Technology Centre for Doctoral Training, University of Warwick, Coventry, CV4 7AL, United Kingdom}

\date{\today}

\begin{abstract} 
Nanodiamonds containing negatively charged nitrogen vacancy centers (${\text{NV}}^{-}$) have applications as localized sensors in biological materials and have been proposed as a platform to probe the macroscopic limits of spatial superposition and the quantum nature of gravity. A key requirement for these applications is to obtain nanodiamonds containing ${\text{NV}}^{-}$ with long spin coherence times. Using milling to fabricate nanodiamonds processes the full 3D volume of the bulk material at once, unlike etching pillars, but has, up to now, limited ${\text{NV}}^{-}$ spin coherence times. Here, we use natural isotopic abundance nanodiamonds produced by ${\text{Si}}_{3}{\text{N}}_{4}$ ball milling of chemical vapor deposition grown bulk diamond with an average single substitutional nitrogen concentration of $121 ~\text{ppb}$. We show that the electron spin coherence times of ${\text{NV}}^{-}$ centers in these nanodiamonds can exceed $\SI{400}{\micro\second}$ at room temperature with dynamical decoupling. Scanning electron microscopy provides images of the specific nanodiamonds containing ${\text{NV}}^{-}$ for which a spin coherence time was measured.
\end{abstract}

\maketitle

\section{Introduction}\label{intro}
The negatively charged nitrogen vacancy center (${\text{NV}}^{-}$) in diamond \cite{doherty_2013} has attracted attention as a tool in quantum information \cite{robledo_2011, bradley_2019}, magnetometry \cite{rondin_2014, barry_2020}, electrometry \cite{dolde_2011, dolde_2014, karaveli_2016, chen_2017b}, and thermometry \cite{acosta_2010, toyli_2013, neumann_2013, plakhotnik_2014} using optically detected magnetic resonance (ODMR). This leverages the optical initialization and readout of the electron spin state of the ${\text{NV}}^{-}$ center, along with the microwave resonance of the spin state transitions, to control the state of the ${\text{NV}}^{-}$ center \cite{jelezko_2006}. In nanodiamonds, the ${\text{NV}}^{-}$ has potential applications in sensing within biological materials as living cells can take in nanodiamonds and remain functional, allowing local sensing within cells \cite{yu_2005, fu_2007, neugart_2007, chang_2008, mcGuinness_2011, schroeder_2012, leSage_2013, kucsko_2013, chipaux_2018, wang_2019, choi_2020, fujiwara_2020}. Also, nanodiamonds containing ${\text{NV}}^{-}$ have been proposed as a platform to probe macroscopic spatial superpositions \cite{scala_2013, yin_2013, wan_2016b, wan_2016a, pedernales_2020, wood_2022} and the quantum nature of gravity \cite{albrecht_2014, bose_2017, marletto_2017}. These proposals require macroscopic spatial superposition states of the nanodiamonds involved, therefore, diamonds with a diameter on the order of \SI{1}{\micro\metre} containing a single ${\text{NV}}^{-}$ center are proposed. Along with large nanodiamonds, the electron spin coherence time, ${T}_{2}$, of the ${\text{NV}}^{-}$ is a critical factor for these experiments. Dynamical decoupling techniques are used to suppress the dephasing of the ${\text{NV}}^{-}$ spin state due to static or slowly changing fluctuations in the environment, maximizing the ${T}_{2}$ time.   

In bulk diamond, ${\text{NV}}^{-}$ ${T}_{2}$ times exceeding $\SI{1}{\second}$ have been observed, using dynamical decoupling, at cryogenic temperatures \cite{abobeih_2018, bar-gill_2013}. At room temperature the longest ${\text{NV}}^{-}$ ${T}_{2}$ time is around $\SI{2}{\milli\second}$, using ${}^{12}\text{C}$ purification and dynamical decoupling \cite{balasubramanian_2009, bar-gill_2013}. However, observed $T_{2}$ times in nanodiamonds are significantly shorter. The longest reported ${T}_{2}$ in micro- or nanodiamonds is $\SI{708}{\micro\second}$ with dynamical decoupling and using isotopically pure ${}^{12}\text{C}$ diamond material that is etched into pillars of diameters $300$ to $\SI{500}{\nano\metre}$ and lengths $\SI{500}{\nano\metre}$ to $\SI{2}{\micro\metre}$ \cite{andrich_2014}. For natural abundance ${}^{13}\text{C}$ micro- or nanodiamonds, the longest ${T}_{2}$ time reported for particles fabricated using etching techniques is $\SI{210}{\micro\second}$ \cite{trusheim_2014} and by milling $\SI{67}{\micro\second}$ \cite{knowles_2014}. 

The ${T}_{2}$ time is sensitive to the dynamics of spins surrounding the ${\text{NV}}^{-}$, hence the shorter times for nanodiamonds containing uncontrolled ${}^{13}\text{C}$ spins. Therefore, it has been suggested that the suppression of ${\text{NV}}^{-}$ ${T}_{2}$ in nanodiamonds is due to defects at the surface \cite{rondin_2010, sangtawesin_2019, guillebon_2020}.  

Here, we show that chemical vapor deposition (CVD) grown diamond with natural ${}^{13}\text{C}$ abundance and nitrogen concentration $< 1 ~\text{ppm}$, can be processed by milling to fabricate nanodiamonds containing ${\text{NV}}^{-}$, with ${T}_{2}$ exceeding $\SI{400}{\micro\second}$ at room temperature. Milling conveniently permits the creation of nanodiamonds from the full 3D volume of the bulk material at once, unlike etching. The nanodiamond ${T}_{2}$ measurements were carried out using confocal fluorescence microscopy (CFM), and the same nanodiamonds were viewed by scanning electron microscopy (SEM).   


\begin{figure*}
	\includegraphics[width=\linewidth]{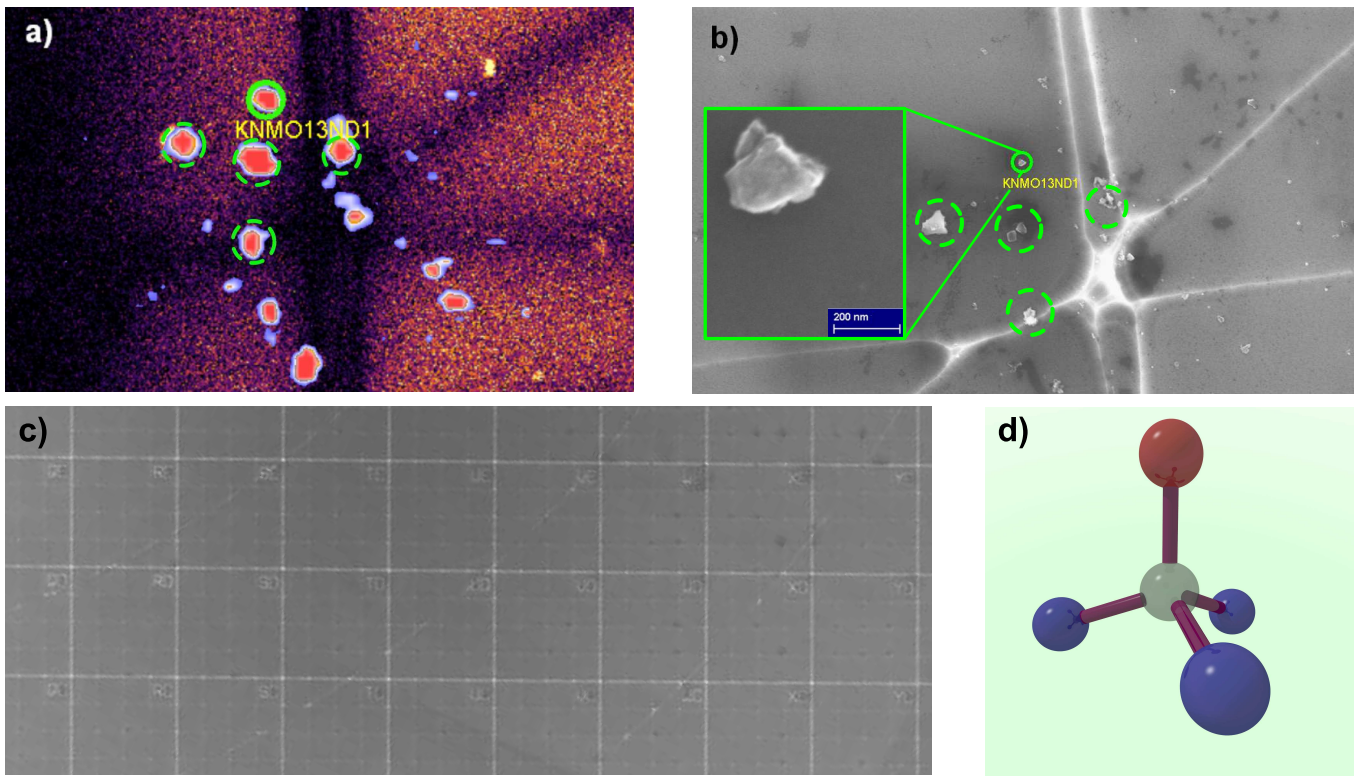}
	\caption{(a) Confocal fluorescence microscopy (CFM) image, in red and blue (lighter grayscale), of nanodiamond KNMO13ND1 (ND1) overlaid onto a reflection image, in purple and orange (darker grayscale), of the grid-marked silicon. Image plotted using the Qudi software suite \cite{binder_2017}. ND1 is identified by the solid green (gray) ring. The dashed green (gray) rings around nearby features are overlaid on both (a) and (b) which, along with the grid, verify that the CFM and scanning electron microscopy (SEM) are both viewing the same nanodiamonds. (b) SEM imaging of the same area in (a) is necessary to provide a measurement of size. The inset shows a higher magnification image of ND1. (c) A lower magnification SEM image of an etched silicon map before nanodiamonds are introduced. Vertical and horizontal grid lines are visible along with arcs that are centered on the center of the silicon map. The grid lines visible in (a), (b) are the intersection of horizontal, vertical, and arc markings. In (a), the grid lines are the dark lines to the right of ND1, and in (b) the grid lines are the bright lines to the right of ND1. Each small square has dimensions $25 ~\text{x} ~\SI{25}{\micro\metre}$. (d) ${\text{NV}}^{-}$ center schematic, three carbon atoms shown in blue (dark gray) at the bottom, one nitrogen in red (gray) at the top, and the transparent (light grey) central sphere is the vacant lattice site. \label{CFM_SEM}}
\end{figure*}

Single-crystal CVD diamond was manufactured by Element Six with an average single substitutional nitrogen concentration of $\SI{121}{ppb}$ measured by electron paramagnetic resonance \cite{frangeskou_2018}, and a natural abundance of ${}^{13}\text{C}$. The expected grown-in ${\text{NV}}^{-}$ concentration was $\SI{0.4}{ppb}$ \cite{edmonds_2012}. Prior to ${\text{Si}}_{3}{\text{N}}_{4}$ ball milling \cite{gines_2018}, the diamonds used for this research were irradiated with $\SI{4.5}{\mega\eV}$ electrons for one minute and annealed for three hours at $\SI{400}{\degreeCelsius}$, four hours at $\SI{800}{\degreeCelsius}$, and two hours at $\SI{1200}{\degreeCelsius}$, similarly to previous methods \cite{frangeskou_2018, chu_2014}. The irradiation time was chosen such that the expected final ${\text{NV}}^{-}$ concentration was approximately $\SI{1}{ppb}$. Given the atomic density of diamond ($\SI{1.77e23}{\per\centi\metre\cubed}$), it was expected that a nanodiamond containing a single center was around $\SI{230}{\nano\metre}$ in diameter.

Reference \cite{gines_2018} provides a detailed report of the fabrication process, x-ray photoelectron spectroscopy measurements of the surface, and particle-size distributions. In summary, the diamond plates were milled with ${\text{Si}}_{3}{\text{N}}_{4}$ balls to avoid magnetic contaminants from steel ball milling. After milling, the sample was acid cleaned in ${\text{H}}_{3}{\text{PO}}_{4}$ and then cleaned in NaOH, to remove the ${\text{Si}}_{3}{\text{N}}_{4}$ contaminants. This process does not remove all the ${\text{Si}}_{3}{\text{N}}_{4}$ contaminants, as the diamond sample gains mass after milling. The nanodiamonds were then annealed in an air atmosphere, dispersed in water, and centrifuged at a relative centrifugal force of $\SI{40e3}{g}$. The air anneal produces nanodiamonds that have surfaces consisting of C-Si, COOH, C=O, C-O, C=C, and C-C bonds.   

The nanodiamonds were held in a suspension of methanol at a density of approximately $\SI{1}{\milli\gram\per\milli\litre}$ and sprayed for three seconds by a nebulizer (Omron MicroAIR U22) into an upturned vial, ensuring that a high density of nanodiamonds were injected. The nanodiamonds were then allowed to precipitate onto silicon wafers. This was to reduce the coffee-ring effect which was often observed in drop casting and to prevent aggregation, which was found to be prevalent when using direct spray applications. Other methods have also been demonstrated previously for mitigating the coffee-ring effect \cite{hees_2011}. $n$-type silicon wafers doped with $\SI{1e15}{\per\centi\metre\cubed}$ of phosphorus were plasma etched using photolithography to create a grid system for locating individual nanodiamonds. This allows verification that the same nanodiamond is being addressed in both the CFM and SEM measurements.

Under CFM, nanodiamonds containing single ${\text{NV}}^{-}$ centers were identified by Hanbury Brown-Twiss (HBT) measurements. HBT measurements quantify the degree of correlation between photon-detection events for different time delays, ${g}^{(2)}(\tau)$. A single ${\text{NV}}^{-}$ cannot emit two photons simultaneously, therefore ${g}^{(2)}(0) = 0$ is expected. However, background fluorescence generates spurious coincidence events so $0 \leq {g}^{(2)}(0) < 0.5$ indicates a single center. Background counts are not subtracted from the HBT data in this paper.  

Those that displayed ODMR were selected and an external magnetic field aligned to the ${\text{NV}}^{-}$ axis. The magnetic field is generated by a permanent magnet on an arm connected to three motors. Two motors rotate the magnet about perpendicular axes that intersect at the position of the sample. These allow rotation in a sphere around the sample without altering the distance between the magnet and the sample. The final motor linearly alters the distance between the magnet and the sample, without changing the angle. Therefore the angular alignment and magnetic field strength can be varied precisely and independently. To align the magnetic field, the fluorescent count rate is monitored and the angle of the magnet adjusted until the magnet can be brought close to the sample without the count rate decreasing. The count rate decreases in the presence of a misaligned field as the ${m}_{s} = 0, \pm 1$ levels are no longer eigenstates of the system and so the spin states mix \cite{rondin_2014}. ODMR could have been used for the magnetic field alignment, however, we found that monitoring the count rate provided a faster alignment. For aligned ${\text{NV}}^{-}$ centers, the fluorescence intensity remains constant as the magnitude of the magnetic field increases \cite{rondin_2014}. Spin-echo decay experiments were then carried out at room temperature to determine $T_{2}$ times using the Hahn echo, XY8-1, and XY8-4 dynamical decoupling pulse sequences. The sizes of the individual nanodiamonds were measured by SEM, as shown in Fig. \ref{CFM_SEM}.

\section{Results and Discussion}\label{results}

\begin{figure*}
	\includegraphics[width=\linewidth]{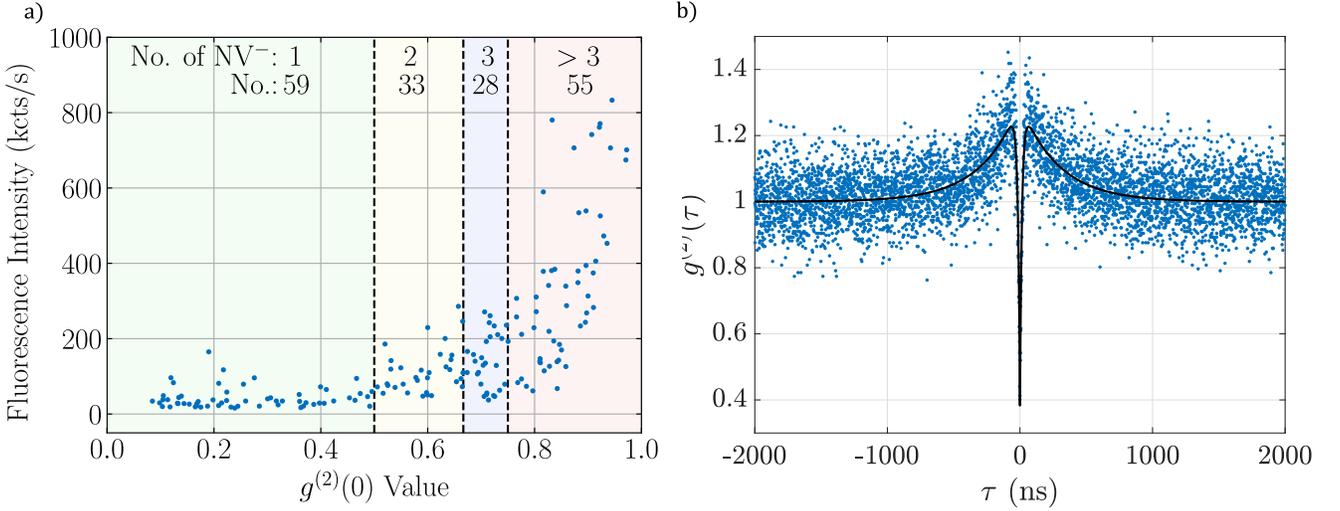}
	\caption{(a) Fluorescence intensity against ${g}^{(2)}(0)$ value. $0 \leq {g}^{(2)}(0) < 0.5$ indicates a single center, $0.5 \leq {g}^{(2)}(0) < 0.67$ indicates two, $0.67 \leq {g}^{(2)}(0) < 0.75$ indicates three, and ${g}^{(2)}(0) \geq 0.75$ indicates more than three. A total of 175 nanodiamonds containing ${\text{NV}}^{-}$(s) were surveyed. (b) An example Hanbury Brown-Twiss (HBT) measurement of photon correlation for ND1. The black curve is a fit to $1-a\left[b \exp\left(-\frac{|\tau|}{c}\right)+\left(1-b\right) \exp\left(-\frac{|\tau|}{d}\right)\right]$, where ${g}^{(2)}(0) = 1 - a$. For this plot, ${g}^{(2)}(0) = 0.38 \pm 0.04$. After multiple measurements, the value quoted in the text of ${g}^{(2)}(0) = 0.39 \pm 0.02$ is reached.\label{g2}}
\end{figure*}

An automated survey collected HBT and fluorescence intensity measurements from 175 nanodiamonds containing ${\text{NV}}^{-}$, as shown in Fig. \ref{g2}(a). Of the nanodiamonds measured, $34\%$ contained a single ${\text{NV}}^{-}$ center whilst $19\%$ contained two, $16\%$ contained three, and $31\%$ contained more than three ${\text{NV}}^{-}$ centers, respectively. We have characterized $0 \leq {g}^{(2)}(0) < 0.5$ to indicate a single center, $0.5 \leq {g}^{(2)}(0) < 0.67$ indicates two, $0.67 \leq {g}^{(2)}(0) < 0.75$ indicates three and ${g}^{(2)}(0) \geq 0.75$ indicates more than three. The survey was conducted at an excitation laser power of $\SI{0.4}{\milli\watt}$.

The sites marked to be included in the HBT survey were identified from their fluorescence under CFM, therefore the number of nanodiamonds that do not contain any ${\text{NV}}^{-}$ centers was not measured. If Poissonian statistics are assumed for the number of nanodiamonds containing one, two, and three ${\text{NV}}^{-}$ centers, a fit of ${e}^{-\lambda}{\lambda}^{x}/{x!}$ gives $\lambda = 1.5 \pm 0.3$. Therefore, it can be estimated that the number of nanodiamonds containing zero ${\text{NV}}^{-}$ in the surveyed region is $40 \pm 12$. However, there are a number of factors that suggest that the data is not well described by Poissonian statistics and that the number of nanodiamonds containing zero ${\text{NV}}^{-}$ centers is an order of magnitude estimate at best. First, while ${\text{NV}}^{-}$ centers may have been incorporated into the bulk diamond material at an approximately constant rate, a range of nanodiamond sizes exist in the sample. Furthermore, selection bias exists in marking sites for surveys. Bright, roughly circular, isolated fluorescence spots are more likely to be identified as a nanodiamond containing ${\text{NV}}^{-}$ and marked for survey than a dim spot, or an extended patch of fluorescence.

Figure \ref{g2}(a) also contains a number of single ${\text{NV}}^{-}$ centers with unusually high fluorescence intensity. We typically see single emitters with $~< 100 \text{kcts/s}$ under CFM, however an number of surveyed sites exceeded this with one approaching $~200 \text{kcts/s}$. A possible explanation for the inflated fluorescence intensity is that the geometry of the milled nanodiamond and the location of the ${\text{NV}}^{-}$ center are, by chance, in the required orientation to act as a waveguide \cite{babinec_2010, momenzadeh_2015}. This could couple more of the emitted fluorescence into the microscope objective than would be the case from a spherical nanodiamond, boosting the measured fluorescence intensity.

 HBT measurements on the nanodiamond labeled KNMO13ND1 (ND1) in Fig. \ref{CFM_SEM} gave the value ${g}^{(2)}(0) = 0.39 \pm 0.02$, as shown in Fig. \ref{g2}(b), indicating that it contained a single ${\text{NV}}^{-}$ center. (For further discussion of this inflated ${g}^{(2)}(0)$, see Appendix \ref{a1}). SEM observations of ND1 were used to estimate that the maximum distance an ${\text{NV}}^{-}$ center could be from the surface (${R}_{\text{max}}$) was $106 \pm 2 ~\si{\nano\metre}$.

\begin{figure*}
	\includegraphics[width=\linewidth]{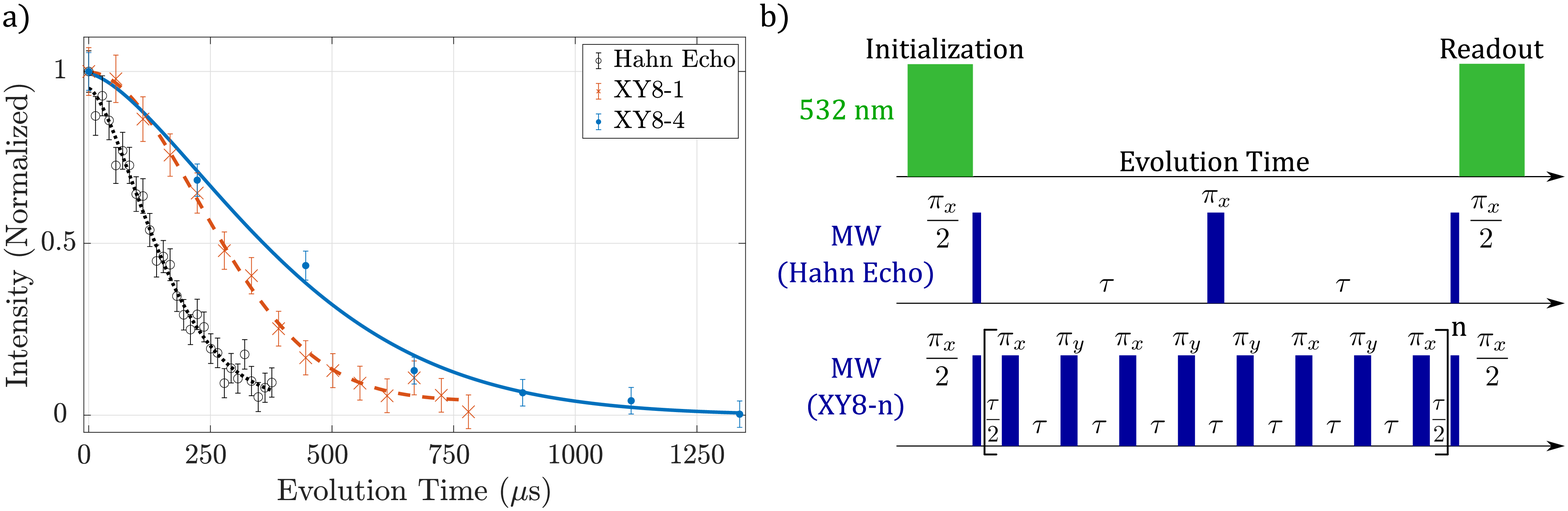}
	\caption{(a) Spin-echo decay measurements of ND1. For the Hahn echo, XY8-1, and XY8-4 pulse sequences, we measure coherence times ${T}_{2}^{\text{HE}} = 177 \pm 24 ~\si{\micro\second}$, ${T}_{2}^{\text{XY8-1}} = 323 \pm 21 ~\si{\micro\second}$, and ${T}_{2}^{\text{XY8-4}} = 462 \pm 130 ~\si{\micro\second}$, respectively. Lines fit by $a+b \exp\left[-{\left(t/{T}_{2}\right)}^{n}\right]$. (b) The $\SI{532}{\nano\metre}$ laser and microwave (MW) pulses applied for the Hahn echo and XY8-n sequences. The bracketed block in XY8-n is repeated n times. \label{T2}}
\end{figure*}

Spin-echo decay measurements were performed on seven nanodiamonds containing a single ${\text{NV}}^{-}$ that also displayed satisfactory ODMR contrast (the difference in fluorescence intensity for an ${\text{NV}}^{-}$ in the ${m}_{s} = 0$ or the ${m}_{s} = \pm 1$ states). The nanodiamond ND1 provided the longest ${T}_{2}$ time of all nanodiamonds measured, as shown in Fig. \ref{T2}(a), with values of ${T}_{2}^{\text{HE}} = 177 \pm 24 ~\si{\micro\second}$, ${T}_{2}^{\text{XY8-1}} = 323 \pm 21 ~\si{\micro\second}$, and ${T}_{2}^{\text{XY8-4}} = 462 \pm 130 ~\si{\micro\second}$ for the Hahn echo, XY8-1, and XY8-4 pulse sequences, respectively. These measurements were taken at an external field strength, measured by ODMR, of $\SI{27}{\milli\tesla}$. Schematics of the pulse sequences are shown in Fig. \ref{T2}(b). 

${T}_{2}$ measurements taken on the same nanodiamond before and after SEM indicated that ${T}_{2}$ was not corrupted. While small variations were observed, this is likely to be due to small changes in the magnetic field alignment \cite{maze_2008, stanwix_2010}. As discussed in Sec. \ref{intro}, the nanodiamond surface is largely oxygen terminated \cite{gines_2018}. Any oxygen based groups that may be removed under SEM will immediately return on venting. The only significant change to the sample after SEM is that background fluorescence increases. We suggest that this is due to electrostatic charging of the silicon under SEM, which then attracts more fluorescent material after venting. This increase in background fluorescence can be seen in Appendix \ref{a1}. 

The measurements in Fig. \ref{T2}(a) were made with the time delay between microwave pulses chosen to sample the peaks of the ${}^{13}\text{C}$ revivals of the spin-echo signal \cite{childress_2006, maze_2008} that is present in ND1 (see Appendix \ref{a2}). This allows the data to be fit by an exponential without sinusoidal components.

\begin{figure*}
	\includegraphics[width=\linewidth]{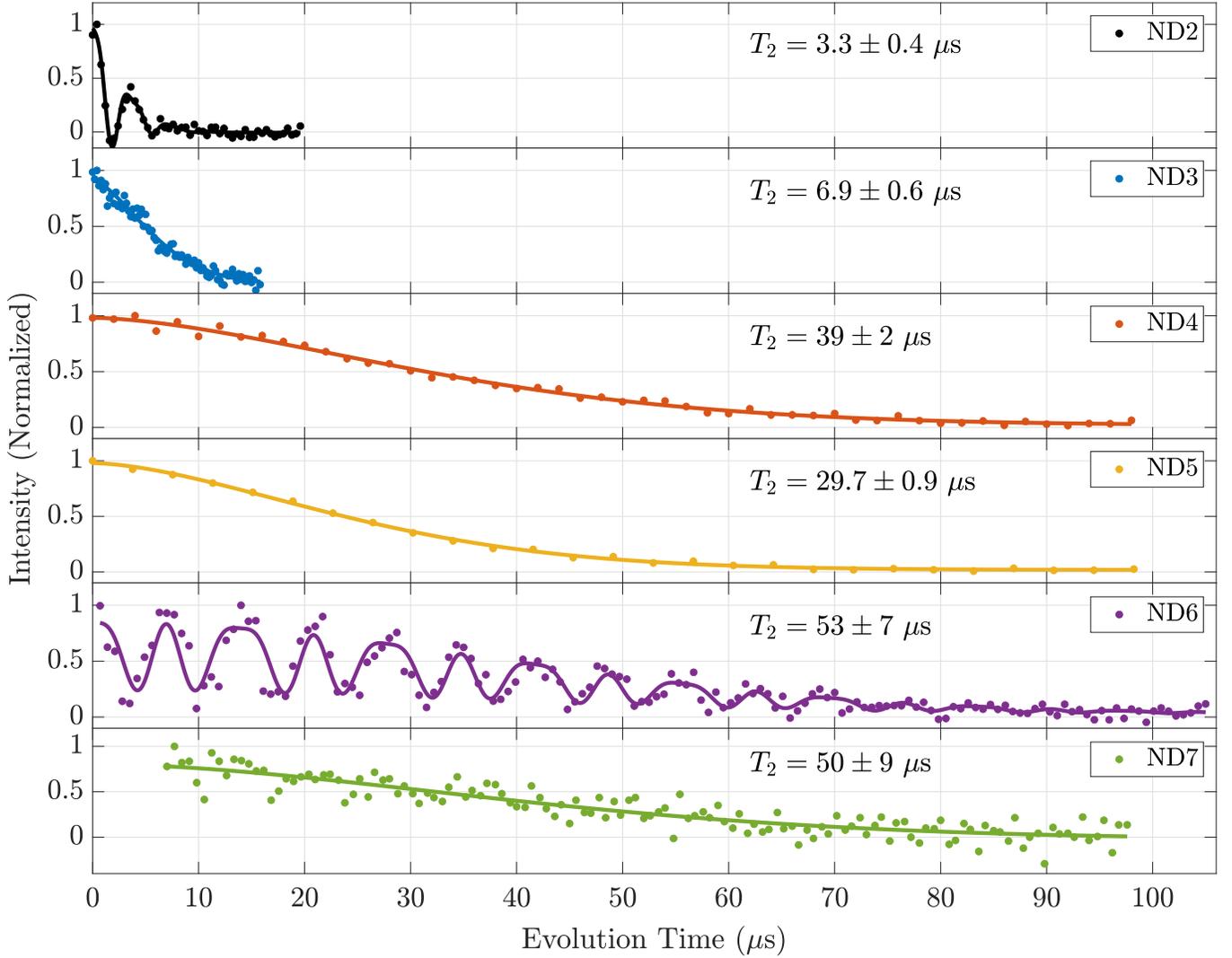}
	\caption{Hahn echo measurements on six nanodiamonds other than ND1 containing single ${\text{NV}}^{-}$. Error bars are not shown as they are smaller than the data points. Dashed lines fit by $a+b \exp\left[-{\left(t/{T}_{2}\right)}^{n}\right]$ or, for those figures that contain ${}^{13}\text{C}$ revivals, $a+b \exp\left[-{\left(t/{T}_{2}\right)}^{n}\right]\left(1-c\ {\sin\left(\frac{\pi t}{d}\right)}^{2}{\sin\left(\frac{\pi t}{g}\right)}^{2}\right)$ \cite{childress_2006, maze_2008}. For the plots with revivals, the periods are as follows: for ND2: $d = 3.6 \pm 0.2 ~\si{\micro\second}$, $g = 4.0 \pm 0.3 ~\si{\micro\second}$, and for ND6: $d = 6.96 \pm 0.05 ~\si{\micro\second}$, $g = 13.9 \pm 0.1 ~\si{\micro\second}$. \label{T2s}}
\end{figure*}

Hahn echo measurements on six other nanodiamonds containing single ${\text{NV}}^{-}$ gave ${T}_{2}^{\text{HE}}$ in the range $3.3$ to $53 ~\si{\micro\second}$, as shown in Fig. \ref{T2s}. The mean ${T}_{2}^{\text{HE}}$ time, including ND1, was $\langle{T}_{2}^{\text{HE}}\rangle = \SI{51}{\micro\second}$. From SEM imaging of this group, including ND1, the mean size was characterized by $\langle{R}_{\text{max}}\rangle = \SI{83}{\nano\metre}$. The six measurements were taken with external magnetic fields, measured by ODMR, that ranged from $26$ to $\SI{50}{\milli\tesla}$.

The ${T}_{2}$ times presented here are, to our knowledge, the longest ${T}_{2}$ times for ${\text{NV}}^{-}$ centers in nanodiamonds with a natural abundance of ${}^{13}\text{C}$, despite these nanodiamonds being produced by milling. Milling produces much larger quantities of nanodiamonds, as it allows for production of nanodiamonds from the full volume of the bulk material, unlike etching. Previous measurements in the literature have reported that milled nanodiamonds contain ${\text{NV}}^{-}$ with ${T}_{2}$ times shorter than those in nanodiamonds produced by etching \cite{knowles_2014, trusheim_2014}. Furthermore, we have introduced a technique with SEM and silicon grid mapping to image the exact nanodiamonds for which we measure ${\text{NV}}^{-}$ spin coherence times. Previous experiments reporting the ${T}_{2}$ time of nanodiamonds did not have a way of getting SEM images of the specific nanodiamond for which they measured spin coherence. Our ${T}_{2}$ time of $\SI{460}{\micro\second}$ is over twice as long as the $\SI{210}{\micro\second}$ reported in Ref. \cite{trusheim_2014} using etching and over six times longer than the $\SI{67}{\micro\second}$ reported in Ref. \cite{knowles_2014} with milling. Five of the six Hahn echo ${T}_{2}$ times in Fig. \ref{T2s} are longer than the longest Hahn echo ${T}_{2}$ times previously reported for milled nanodiamonds of $2$ to $\SI{6}{\micro\second}$ in Ref. \cite{knowles_2014}. It should be noted that the nanodiamond pillars in Ref. \cite{trusheim_2014} are of a similar size to those measured here, with diameter $50 \pm \SI{15}{\nano\metre}$ and height $150 \pm \SI{75}{\nano\metre}$. However, in Ref. \cite{knowles_2014}, the nanodiamonds are smaller, with the majority of diameters within $10$-$\SI{35}{\nano\metre}$. 

The nanodiamonds surveyed for Fig. \ref{g2}(a) were in a different region on the silicon grid to those nanodiamonds that were measured for spin coherence times. We estimate that around $80$ nanodiamonds containing single ${\text{NV}}^{-}$ centers were investigated to measure seven ${T}_{2}$ times. There are a number of reasons why a single ${\text{NV}}^{-}$ may not produce a ${T}_{2}$ measurement. For instance, if the ODMR contrast is too low, the number of measurement repeats required to reach an acceptable signal-to-noise ratio becomes prohibitive. ${\text{NV}}^{-}$s can display low ODMR contrast if they are too far from the wire delivering microwave excitation, and we have also observed that some ${\text{NV}}^{-}$, even if close to the wire, have little to no ODMR contrast. This lack of ODMR contrast has been observed before \cite{vandam_2019}. ${\text{NV}}^{-}$s can also fail to produce a ${T}_{2}$ measurement due to the limited range of motion of the arms that align the magnet to the ${\text{NV}}^{-}$ axis to avoid crashing into the sample stage. If the magnet cannot be aligned to the ${\text{NV}}^{-}$ axis, then a magnetic field cannot be applied to break the degeneracy of the ${m}_{s} = 0, \pm 1$ levels without also significantly degrading the  ${T}_{2}$ time \cite{maze_2008, stanwix_2010}.   

Under the assumption that all the nanodiamonds are spherical, and the ${\text{NV}}^{-}$s are located at the center of the sphere, we would expect that larger nanodiamonds would correlate with longer ${T}_{2}$ times. The further the ${\text{NV}}^{-}$ is from the surface, the decohering effects of the surface are suppressed. However, in practice our milled nanodiamonds are far from spherical, and the ${\text{NV}}^{-}$ center could be anywhere within the volume of the nanodiamond. As such, we do not observe a correlation between the size of the nanodiamond and the ${T}_{2}$ time (see Appendix \ref{a3}).  

\section{Conclusion\label{conc}}
We observed a nanodiamond containing a single ${\text{NV}}^{-}$ electron spin coherence exceeding $\SI{400}{\micro\second}$, with dynamical decoupling. For other nanodiamonds containing single ${\text{NV}}^{-}$ centers, the average ${T}_{2}$ time measured by the Hahn echo sequence across the sample was $\SI{51}{\micro\second}$. All spin coherence measurements were performed at room temperature. The nanodiamonds containing ${\text{NV}}^{-}$ were fabricated from CVD diamond bulk material by ${\text{Si}}_{3}{\text{N}}_{4}$ ball milling \cite{gines_2018, frangeskou_2018}. CVD allows diamond to be grown with low, and controllable, defect concentrations and milling permits the conversion of the entire bulk sample into nanodiamonds quickly, unlike masked etching of pillars. We have also used etched grid markings in silicon to be able to address specific nanodiamonds, that provided ${T}_{2}$ measurements, under SEM.

These ${T}_{2}$ times demonstrate that nanodiamonds produced by milling can contain ${\text{NV}}^{-}$ centers with ${T}_{2}$ times that are comparable with or longer, than those produced by etching. These ${T}_{2}$ times should enable AC magnetometry with a sensitivity on the order of $\SI{100}{\nano\tesla\per\hertz\tothe{1/2}}$ \cite{rondin_2014}. Furthermore, the high-volume fabrication enabled by milling is compatible with applications such as sensing \cite{yu_2005, fu_2007, neugart_2007, chang_2008, mcGuinness_2011, schroeder_2012, leSage_2013, kucsko_2013, chipaux_2018, wang_2019, choi_2020, fujiwara_2020} and nanodiamond levitation \cite{scala_2013, yin_2013, wan_2016b, wan_2016a, pedernales_2020, bose_2017, marletto_2017, hoang_2016b, hsu_2016, pettit_2017, frangeskou_2018, delord_2018, conangla_2018, obrien_2019}. 

\begin{acknowledgements}
G.A.S.'s PhD studentship is funded by the Engineering and Physical Sciences Research Council (EPSRC) Centre for Doctoral Training in Diamond Science and Technology (Grant No. EP/L015315/1). J.E.M.'s PhD studentship is funded by the Royal Society. B.L.G. is supported by the Royal Academy of Engineering. This work is supported by the UK National Quantum Technologies Programme through the NQIT Hub (Networked Quantum Information Technologies), the Quantum Computing and Simulation (QCS) Hub, and the Quantum Technology Hub for Sensors and Metrology with funding from UKRI EPSRC Grants No. EP/M013243/1, No. EP/T001062/1, and No. EP/M013294/1, respectively. G.W.M. is supported by the Royal Society.
\end{acknowledgements}

\appendix

\section{Electron Irradiation\label{ei}}
The irradiation was performed by Synergy Health in Swindon, United Kingdom. The beam is not well characterized, however, the beam current is approximately $\SI{20}{\milli\ampere}$ and produces vacancies at a rate of around 0.3 ppm/hr. Based on prior experience of using the irradiation facility, the one minute exposure time was chosen such that the expected final ${\text{NV}}^{-}$ concentration was approximately $\SI{1}{ppb}$.

\section{ND1 photon autocorrelation\label{a1}}
HBT measurements on ND1 gave a fitted autocorrelation value of ${g}^{(2)}(0) = 0.39 \pm 0.02$. Whilst this satisfies the ${g}^{(2)}(0) <  0.5$ condition for a single ${\text{NV}}^{-}$, it is larger than the values we typically observe for single ${\text{NV}}^{-}$ in nanodiamonds at room temperature. The inflated ${g}^{(2)}(0)$ value could be caused either by ND1 containing two ${\text{NV}}^{-}$ centers with the emission intensity of one suppressed, or ND1 containing a single ${\text{NV}}^{-}$ with a high background count rate, as background counts are not subtracted from the HBT data.

\begin{figure*}
	\includegraphics[width=\linewidth]{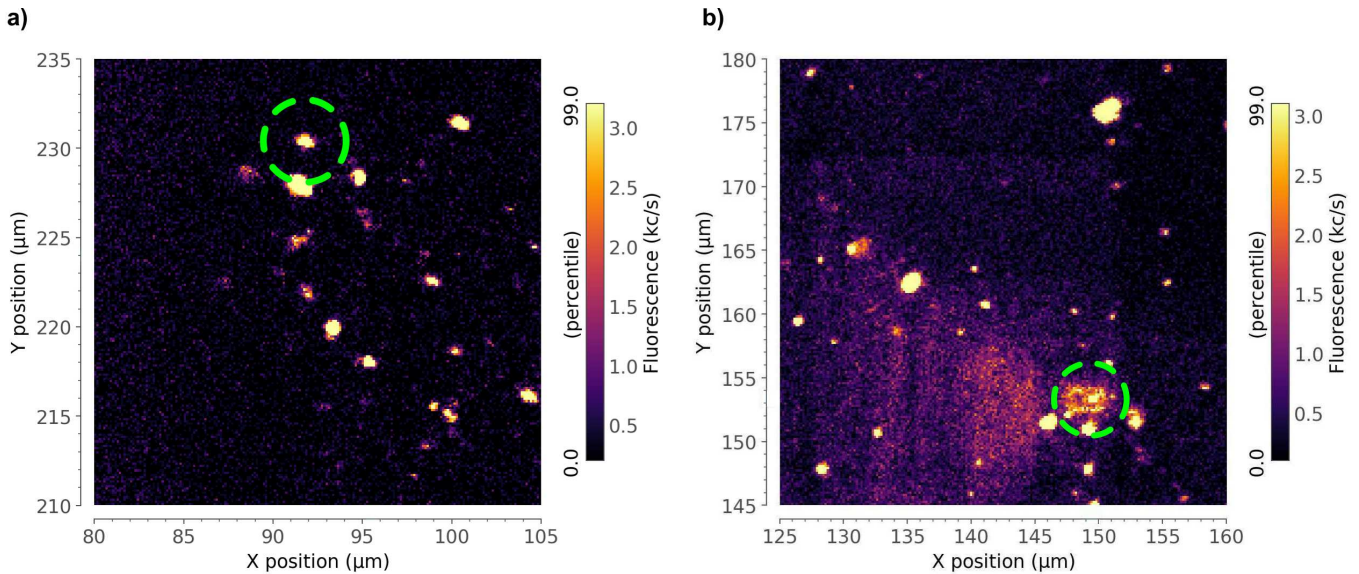}
	\caption{Confocal fluorescence microscopy (CFM) images of ND1 identified by a green (light gray) dashed ring. (a) CFM taken prior to scanning electron microscopy (SEM) observation of ND1. (b) CFM taken after SEM observation of ND1. A bright rectangle of background fluorescence is visible surrounding ND1. This background is not present prior to SEM observation. Images plotted using the Qudi software suite \cite{binder_2017}. \label{conf}}
\end{figure*}

The first potential cause for emission suppression is if the two ${\text{NV}}^{-}$ have different orientations in the diamond lattice then the input polarization of the $\SI{532}{\nano\metre}$ excitation could couple preferentially with one orientation over the other. However, a number of experimental observations suggest that if there are two ${\text{NV}}^{-}$ then they have the same orientation. First, there is only one pair of peaks observed in ODMR when an external magnetic field is applied. Second, a magnet can be aligned and moved from a distance of approximately $\SI{30}{\milli\metre}$ to approximately $\SI{5}{\milli\metre}$ from ND1 without changing the emitted fluorescence intensity. If there were two ${\text{NV}}^{-}$ of different orientations, the off-axis magnetic field would further suppress the counts from one of them. Finally, multiple HBT measurements were taken with and without an aligned magnetic field and there was no clear difference in the ${g}^{(2)}(0)$ value between the two cases. Once again, if there were two ${\text{NV}}^{-}$ of different orientations, the magnetic field should change the level of suppression of the emission from one of the ${\text{NV}}^{-}$, changing the ${g}^{(2)}(0)$ value.

Second, if one of the two ${\text{NV}}^{-}$ is charge switching to ${\text{NV}}^{0}$, then its average fluorescence intensity is reduced. However, step changes in count rate due to charge-state switching have never been observed for ND1. It could be that charge switching is happening at a high frequency that cannot be seen as step changes in the fluorescence count rate, however, the charge switching would have had to have been consistently high frequency over the many hours of ND1 observations.

Finally, high background count levels can inflate ${g}^{(2)}(0)$ values by increasing the number of coincident counts. For ND1, dynamical decoupling measurements were taken after the sample had been observed by SEM. After SEM, a rectangle of background fluorescence, more intense than the global background, was visible around ND1, as shown in Fig. \ref{conf}. It is possible that the electron dose incident on the silicon on which ND1 sits caused fluorescent material to electrostatically stick to the silicon around ND1. Furthermore, a HBT measurement taken prior to SEM observation gave a value of ${g}^{(2)}(0) = 0.21 \pm 0.11$. 

Therefore, despite the inflated value for an ideal single ${\text{NV}}^{-}$ center, ${g}^{(2)}(0) = 0.39 \pm 0.02$, secondary observations and the ${g}^{(2)}(0) <  0.5$ condition being satisfied suggest that ND1 is a nanodiamond containing a single ${\text{NV}}^{-}$ in the presence of a high level of background counts.

\section{${}^{13}\text{C}$ revivals\label{a2}}

\begin{figure*}
	\includegraphics[width=0.85\linewidth]{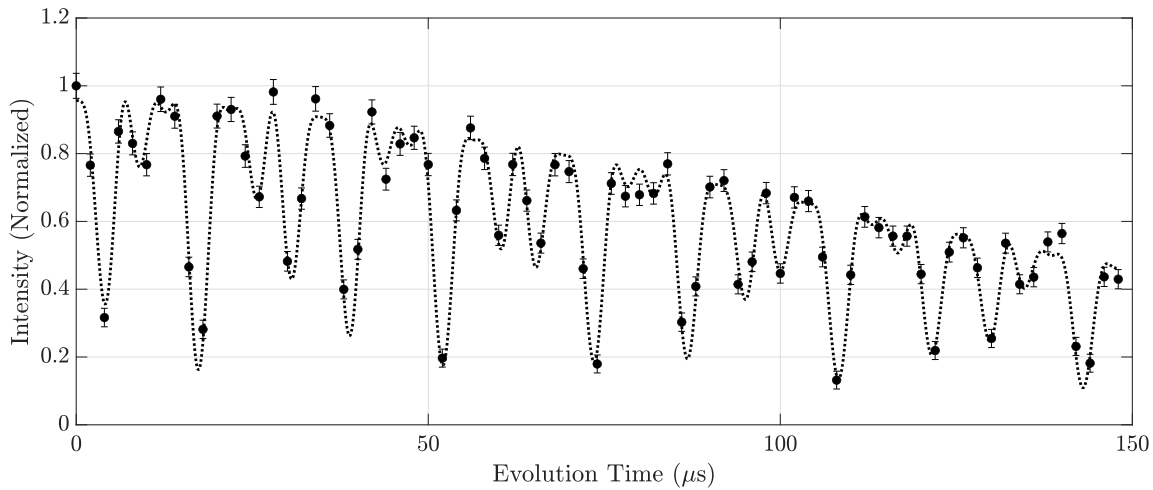}
	\caption{Hahn echo measurement of ND1 with shorter inter-point spacing than in Fig. \ref{T2} of the main text. Revivals in spin-echo signal due to ${}^{13}\text{C}$ spins are well fit by $a+b \exp\left[-{\left(t/{T}_{2}\right)}^{n}\right]\left(1-c\ {\sin\left(\frac{\pi t}{d}\right)}^{2}{\sin\left(\frac{\pi t}{g}\right)}^{2}\right)$ \cite{childress_2006, maze_2008}, where $d = 6.975 \pm 0.007 ~\si{\micro\second}$ and $g = 11.44 \pm 0.02 ~\si{\micro\second}$. \label{rev}}
\end{figure*}

Dynamical decoupling sequences, such as Hahn echo and XY8-n, act to cancel magnetic fluctuations local to the ${\text{NV}}^{-}$ center that are static, or change slowly with respect to the $\pi$ pulse spacing. However, a prominent dynamical change in the magnetic environment is due to the precession of ${}^{13}\text{C}$ spins. Slight differences in the precession frequency of ${}^{13}\text{C}$ spins in the surrounding spin bath due to slight misalignment between the external magnetic field and the ${\text{NV}}^{-}$ axis and hyperfine interaction between those spins and the ${\text{NV}}^{-}$ spin induces decoherence \cite{stanwix_2010}.

Alongside the decoherence effects of the ${}^{13}\text{C}$ spin bath, individual ${}^{13}\text{C}$ spins close to the ${\text{NV}}^{-}$ can couple coherently through the hyperfine interaction causing collapses and revivals in the fluorescence intensity in spin-echo measurements \cite{childress_2006, maze_2008}. Examples of these collapses and revivals are shown in Fig. \ref{rev} in a Hahn echo measurement on ND1, where the oscillation is well described by the interaction between the ${\text{NV}}^{-}$ and one proximal ${}^{13}\text{C}$ spin.

The revivals are not present in Fig. \ref{T2} of the main text as the sampled evolution times are chosen to match the peaks of the revivals in the spin-echo signal. This allows the measurement to be run with far fewer data points than would be required to adequately fit the oscillations across the full $0 - \SI{400}{\micro\second}$ range. The exponential envelope, and as such the ${T}_{2}$ times, can then be fit with a higher signal-to-noise ratio for the same elapsed time as each data point can be repeated a greater number of times. 

For the longer pulses sequences, XY8-1 and XY8-4, the time required to collect enough data points with a good enough signal-to-noise ratio to properly fit the revivals becomes prohibitive. Therefore, the sampled evolution times are chosen to match the revival peaks once again. However, the minimum interpulse wait time, $\frac{\tau}{2}$, in Fig. \ref{T2}(b) of the main text must be an integer multiple of the revival time. This integer requirement to hit the revival peaks means that the total evolution time in XY8-4 steps in integer multiples of 32 times the fitted revival period in Fig. \ref{T2}. This is why there are only seven points in the XY8-4 measurement in Fig. \ref{T2}(a) of the main text, as that is the most tightly spaced evolution time sampling possible whilst still hitting revival peaks.    

\begin{figure}
    \centering
	\includegraphics[width=\linewidth]{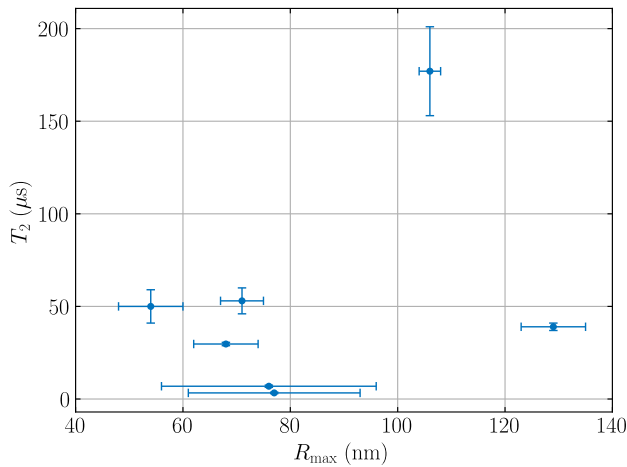}
	\caption{Spin echo ${T}_{2}$ times against the estimated maximum distance the ${\text{NV}}^{-}$ could be from the surface (${R}_{\text{max}}$), plotted for all seven nanodiamonds that contained an ${\text{NV}}^{-}$ for which a ${T}_{2}$ was measured. \label{size_T2}}
\end{figure}

\section{Nanodiamond size vs. ${T}_{2}$ time\label{a3}}

Using the grid markings etched onto the silicon the nanodiamonds are deposited on, each nanodiamond that contained an ${\text{NV}}^{-}$ for which a ${T}_{2}$ was measured was viewed under SEM. Figure \ref{size_T2} shows that we did not observe a correlation between nanodiamond size and ${T}_{2}$ time.

There are limitations to characterizing the nanodiamond size under SEM. The two-dimensional image allows the projected size to be measured, but provides no information as to the depth of the nanodiamond. Nanodiamonds are sometimes deposited in clumps, leaving it unclear under SEM as to whether there is one large nanodiamond that contains the ${\text{NV}}^{-}$, or a smaller nanodiamond containing the ${\text{NV}}^{-}$ next to another small nanodiamond that does not contain an ${\text{NV}}^{-}$. This is the case for the two data points in Fig. \ref{size_T2} that have ${T}_{2} < \SI{10}{\micro\second}$.

Even if the SEM observations could provide perfect information on the size and shape of each nanodiamond, we have no knowledge of the location of the ${\text{NV}}^{-}$ center within the diamond. Therefore, ${R}_{\text{max}}$ is an estimate from the SEM images of the maximum distance that the ${\text{NV}}^{-}$ center could be from the surface. These factors, along with those discussed in the main text, all contribute to mask any possible correlation between the nanodiamond size and ${T}_{2}$ time.

\bibliography{bib.bib} 

\end{document}